\documentclass[11pt,a4paper]{article} 
\usepackage{mathptmx}
\usepackage{graphicx}
\usepackage{blindtext}
\usepackage{amsmath}
\usepackage{sectsty}

\allsectionsfont{\centering}

\title{\bf A COMPARISON OF CONSTITUTIVE MODELS OF BLOOD}

\author{CORINA S. DRAPACA\thanks{Corresponding Author}\\ Pennsyvania State University, University Park, PA 16802, USA\\ ZHIFENG ZHANG\\ Pennsyvania State University, University Park, PA 16802, USA\\
and\\ RUIFAN MENG\\ Hohai University, Nanjing 210000, Jiangsu Province, China
}

\begin{document}
\maketitle

\begin{abstract}
{\fontsize{8pt}{12pt}\selectfont
Mathematical models that accurately predict the mechanical behavior of blood can contribute to the development of biomedical devices and medications which are relevant in clinical applications. The models existing in the literature are complex enough in order to agree with various {\it in vitro} experimental observations. Latest technological advancements opened the possibility of studying blood flow {\it in vivo} which could play an important role in the creation of biocompatible implants for health monitoring and treatment purposes. However, most of existing models may fail to predict blood behavior {\it in vivo} because they require many parameters which are difficult to find {\it in vivo} and they do not incorporate pertinent coupled chemo-mechanical dynamics specific to blood flow in a living body. Recently, we used the fractional model of continuum mechanics proposed by Drapaca and Sivaloganathan [J. Elast.,107: 105-123,  2012] to study blood circulation. In this mathematical formulation the spatial derivatives of the rate of deformation tensor are expressed using Caputo fractional derivatives. The aim of this paper is to compare the Poiseuille flows of blood through an axi-symmetric circular rigid and impermeable pipe where the blood is described by the above mentioned fractional model, the Casson's model, and the power law model. Although the velocity profiles of these three models look similar, the fractional model provides a better fitting to published experimental data.  
}
\end{abstract}

\hspace*{0.35cm}{\bf AMS 2010 Subject Classification:} 45K05, 76A05, 76Z05, 92C10, 92C35\\ 
\hspace*{0.85cm} {\bf Keywords:} Fractional Calculus, Integro-Differential Equations, Blood Models, Poiseuille Flow


\section{\bf INTRODUCTION}

Blood is a complex fluid mixture of components that actively interact chemically and mechanically not only among themselves but also with the vessels containing them. While the mechanics dominates the flow of blood through large vessels where the blood behaves as a viscous Newtonian fluid, in the smaller vessels the coupled chemo-mechanical dynamics become important and the many non-Newtonian models proposed in the literature highlight the inherent difficulties encountered in properly describing the mechanisms of blood flow at smaller length scales. Most of these mathematical models have numerous parameters that may be found from {\it in vitro} experiments \cite{apostol,bess,ku,pop,pri,srir,yil}. However, finding these parameters {\it in vivo} may be very difficult if not impossible due to the very diverse entangled chemo-mechanical dynamics of blood and its surroundings existing in a living body \cite{dav,iad}. In addition, given today’s efforts in building implantable biomedical devices capable to accurately monitor the health state of humans and ultimately interact with the observed living process such as blood flow, using models whose parameters can be found only in the lab could not only inhibit technological progress but also create devices that could be potentially dangerous in medical applications. \\

Finding parameters that reliably describe the complex blood flow {\it in vivo} (aka biomarkers) requires new mathematical models. In \cite{west}, West noticed that {\it ``Understanding complexity [...] requires a new way of modeling and consequently more innovative thinking. [...] Such a perspective might well be provided by the fractional calculus that is able to quantify the coupling of variations in phenomena across widely separated scales in both space and time.''} Fractional calculus, a centuries-old field of mathematical analysis
~\cite{leibniz,liouville,riemann}, uses integro-differential operators whose kernels are power functions of fractional order. Thanks to their integro-differential representations, 
fractional order derivatives have been successfully used in numerous applications  to model stochastic, multiple (possibly entangled) time and/or length scales and non-local phenomena in numerous physical systems ~\cite{samko,pod,hilfer,tar,old_spa,bal,mili,west}. In particular, non-local models of kinematics based on fractional calculus have been proposed to address the local nature of continuum mechanics that makes this theory difficult to use in studies of dislocations and fracture in materials, mechanical behavior of liquid crystals and biological tissues where the stresses and body forces at a point could depend on the thermodynamic state of the entire body ~\cite{carp1,carp2,cott,di1,di2,os,sum1,sum2,sum3}. A generalization of the classical continuum mechanics framework that uses fractional derivatives to represent both the non-local kinematics and non-local stresses corresponding to contact forces with long ranges was given by Drapaca and Sivaloganathan in \cite{dra_siv}. These models of fractional continuum mechanics are supported by the fractional differential geometry on manifolds and corresponding fractional differential forms develeoped by ~\cite{albu1,albu2,cot}, while the physical and geometric interpretations of the fractional order kinematic structure could be inferred from the physical and geometric interpretations of the fractional order integrals and derivatives given in ~\cite{pod1}. \\

Recently, we applied the fractional model of continuum mechanics proposed in \cite{dra_siv} to blood flow \cite{dra2018}. In this formulation the Cauchy stress tensor depends linearly on a generalized rate of deformation tensor whose representation involves Caputo fractional order spatial derivatives of order $\alpha\in {\bf R}$. The physical parameter $\alpha$ is assumed to be a measure of the long range interactions among fluid's particles during flow and could vary with time, location, temperature, pressure, shear rate, and/or concentration of particles. The second physical parameter of this model is the constant of proportionality $\mu$ that relates the stress tensor and this generalized rate of deformation tensor. Parameter $\mu$ depends on a characteristic length scale specific to the problem under investigation and the fractional order $\alpha$. When $\alpha=1$ parameter $\mu$ reduces to the apparent viscosity of the fluid. The fundamental difference between this fractional model and the numerous non-Newtonian viscosity models of blood existing in the literature is as follows. The fractional model uses non-local integro-differential operators in the representation of the rate of deformation tensor which {\it entangle} the aggregation of particles and the chemo-mechanical deformation of the complex fluid (blood). On the other hand, in the non-Newtonian models the stress tensor is decomposed in a product between the rate of deformation tensor and an apparent viscosity that may depend on time, temperature, mechanical invariants, or concentrations of fluid's components. Such a representation of the stress tensor does not account for the chemo-mechanical coupling present during blood flow {\it in vivo} because the rate of deformation tensor is separated from the mechano-chemical interactions of the blood's components and therefore the chemically-driven part of the flow is not accounted for. In addition, the rate of deformation tensor involves spatial differential operators which is a local representation and thus any possible non-local effects are usually hidden in some extra ad-hoc parameters added to the model in order to explain experimental observations.\\

In this paper we consider the fully developed steady flow of an incompressible non-Newtonian fluid due to an externally imposed pressure gradient and in the absence of body forces through a circular pipe. The walls of the pipe are assumend to be rigid and impermeable, and a no slip boundary condition is imposed at the structure-fluid interface. The aim of the paper is to compare the solutions of the Poiseuille flow that correspond to the following three constitutive models of blood: the fractional model discussed above, the power-law model, and the Casson's model. Each of these models has two parameters. The power-law and Casson's constitutive equations are chosen for the comparrison because these laws and some variations of them have been used by \cite{john,nadim,sou,fung} to predict blood's velocity profiles which agree with some experimental observations \cite{apostol,park,santi,zhong}. The Casson's model involves a yielding condition to describe viscoplasticity of blood observed {\it in vitro}. However, the power law and Casson's models have some unphysical features. The power law constitutive equation is characterized by an unbounded viscosity function and a zero viscosity at zero shear rate, while Casson's apparent viscosity is singular and unbounded at zero shear rate. On the other hand, generalizations of these two models that do not suffer from the above mentioned limitations have more physical parameters which are difficult to find experimentally {\it in vivo}. The fractional model does not have these unphysical characteristics. Our comparison shows that the fractional model produces similar velocity profiles to the ones generated by the power law and Casson's models. In addition, best fit curves to experimental data taken from \cite{long} suggest that the fractional model provides the better fit among the three models.\\

The structure of the paper is as follows. In the next section we review the constitutive equations for the fractional, power law and Casson models. In section 3 we present the analytic solutions to the Poiseuille flow in a circular pipe for these three models, while in section 4 we show some numerical simulations of the velocity profiles and curve fitting results. The paper ends with a section of conclusions and future work.   

\section{Constitutive Models}

In this section we provide a brief presentation of the fractional, power law, and Casson models.

\subsection{Fractional Model}

The following definitions are taken from \cite{dra2018}.\\

{\bf Definition 1:} Let $\Omega$ and $\Omega_{t},\,t>0$ be two open subsets of ${\bf R}^{3}$. Let $\pmb{\alpha}$ be a 3x3 matrix whose elements $\alpha_{Ii}(t),\,I,i=1,2,3$ are continuous functions of $t > 0$ such that $\forall t>0,\,I,i=1,2,3$ either $-\infty<\alpha_{Ii}(t)<0$ or $m-1<\alpha_{Ii}(t)\leq m,\,m=1,2,3...$, and let $\pmb{\chi}(\cdot;\,t,\,\pmb{\alpha}(t)):\Omega \rightarrow \pmb{\chi}(\Omega;\,t,\,\pmb{\alpha}(t))\equiv\Omega_{t,\pmb{\alpha}(t)}$ be a family of functions in $L^{1}(\Omega)$. The {\it deformation of order} $\pmb{\alpha}(t)$ of a body occupying region $\Omega$  at $t=0$ and region $\Omega_{t,\,\pmb{\alpha}(t)}$ at time $t>0$ is determined by the position $\bf{x}$ of the material points in space as a function of the reference position ${\bf X}$ at $t=0$, time $t>0$ and $\pmb{\alpha}(t)$ which is given by:
\begin{enumerate}
\item[$\bullet$] if $\alpha_{Ii}(t)=m,\,\forall t>0,\;I,i=1,2,3,\,m=1,2,3,...$ and $\pmb{\chi}(\cdot;\,t,\,\pmb{\alpha}(t))$ is in $\mathcal{C}^{m+1}(\Omega)$:
\begin{equation}
x^{i}=\displaystyle \frac{\partial^{m-1}\chi^{i}({\bf X},t)}{\partial (X^{I})^{m-1}}
\label{def1_eq1}
\end{equation}
\item[$\bullet$] if $-\infty<\alpha_{Ii}(t)<0,\,\forall t>0,\,I,i=1,2,3$
\begin{align}
x^{i}=&\displaystyle \frac{1}{\Gamma(-\alpha_{1i}(t))\Gamma(-\alpha_{2i}(t))\Gamma(-\alpha_{3i}(t))} \nonumber\\
       \times & \iiint_H \frac{\chi^{i}({\bf Y},t) dY^{1}dY^{2}dY^{3}}{|X^{1}-Y^{1}|^{1+\alpha_{1i}(t)}
       |X^{2}-Y^{2}|^{1+\alpha_{2i}(t)}|X^{3}-Y^{3}|^{1+\alpha_{3i}(t)}},
\label{def1_eq2}
\end{align}
\item[$\bullet$] if $m-1<\alpha_{Ii}(t) < m,\,\forall t>0,\; I,i=1,2,3,\; m=1,2,3....$:
\begin{align}
x^{i}=&\displaystyle \frac{1}{\Gamma(m-\alpha_{1i}(t))\Gamma(m-\alpha_{2i}(t))\Gamma(m-\alpha_{3i}(t))} \nonumber\\
       \times & \frac{\partial^{m-1}}{\partial (X^{I})^{m-1}} \iiint_H \frac{\chi^{i}({\bf Y},t) dY^{1}dY^{2}dY^{3}}{|X^{1}-Y^{1}|^{1-m+\alpha_{1i}(t)}
       |X^{2}-Y^{2}|^{1-m+\alpha_{2i}(t)}|X^{3}-Y^{3}|^{1-m+\alpha_{3i}(t)}},
\label{def1_eq3}
\end{align}
\end{enumerate}
In addition, $\pmb{\chi}$ is assumed to be zero on the boundary and outside the region of integration $H$. When $m=1$, equation (\ref{def1_eq1}) becomes the mathematical representation of deformation known from the classical theory of continuum mechanics. In the above equations $\displaystyle \Gamma(s)=\int_{0}^{\infty} t^{s-1} \exp(-t)dt$ is the gamma function. \\

The region of integration $H=[a_1,\,b_1]\times [a_2,\,b_2]\times [a_3,\,b_3]\subset {\bf R}^{3}$ is considered to be the {\it region of influence} of ${\bf X}$ that contains all the material points involved in the deformation of ${\bf X}$ into ${\bf x}$ \cite{dra_siv,dra2018}. The limits $a_i,\,b_i,\,i=1,2,3$ may be functions of ${\bf X},\,t$ and $\pmb{\alpha}$. $H$ is determined by the physics of interactions between particles, by the geometry of the domain occupied by the body under observation, or by curve fitting to experimental data. The components of matrix $\pmb{\alpha}(t)$ could be envisioned as measures of the dynamic (chemo-mechanical) deformations of a body with evolving microstructure. \\

{\bf Definition 2:} The {\it deformation gradient of order} $\pmb{\alpha}(t)$, with $t>0$ and $\alpha_{Ii}\in {\bf R},\,I,i=1,2,3$ is either:
 \begin{equation}
 {\bf F}_{\pmb{\alpha}(t)}=\displaystyle \left(\frac{\partial x^{k}}{\partial X^{K}} \right)_{K,k=1,2,3},\;
{\rm if}\; \alpha_{Ii}(t)>0,\,\forall t>0,\,I,i=1,2,3,
 \label{def2_eq1}
 \end{equation}
 or:
 \begin{equation}
 {\bf F}_{\pmb{\alpha}(t)}={\bf x},\;{\rm if}\; -\infty < \alpha_{Ii}(t)<0,\,\forall t>0, \,I,i=1,2,3.
 \label{def2_eq2}
 \end{equation}
 
We notice that the deformation gradient of order $\pmb{\alpha}(t)$ given by formulas (\ref{def2_eq1}) and (\ref{def2_eq2}) resembles the definitions of the left-sided and right-sided Riemann-Liouville fractional derivatives of order $\alpha_{Ii}(t),\,I,i=1,2,3$. Using some properties of fractional order derivatives it can be shown that inversions can be defined only for the deformations (\ref{def1_eq1}) and (\ref{def1_eq3}). The very long spatial memory of each material point during a deformation (\ref{def1_eq2}) is lost and thus an inversion of this motion is not possible. \\

Other strain and strain rate tensors of order $\pmb{\alpha}(t)$ can be further obtained by combining definitions known from continuum mechanics and formulas (\ref{def2_eq1}) or (\ref{def2_eq2}). For instance, the {\it velocity gradient tensor of order} $\pmb{\alpha}(t)$ is given by:
\[
{\bf L}_{\pmb{\alpha}(t)}=\displaystyle \left( \frac{\partial v^{k}}{\partial x^{k}} \right)_{k=1,2,3}
\]
where ${\bf v}$ is the velocity field associated to the deformation given in definition 1. It follows that the rate of deformation tensor of order $\pmb{\alpha}(t)$ is ${\bf D}_{\pmb{\alpha}(t)}=\left( {\bf L}_{\pmb{\alpha(t)}} + {\bf L}_{\pmb{\alpha(t)}}^{T} \right)/2$.  
The constitutive law for an incompressible fluid is now:
\begin{equation}
\displaystyle {\pmb{\sigma}}=-p{\bf I}+2\mu {\bf D}_{\pmb{\alpha}(t)}
\label{eq3}
\end{equation}
where ${\pmb{\sigma}}$ is the Cauchy stress tensor, $p$ is the hydrostatic pressure, and $\mu$ is a physical parameter of the model.

In particular, if $H$ is independent of $t$ and $\pmb{\alpha}(t)$, and $\pmb{\alpha}$ is a constant matrix, then the velocity fields $\displaystyle {\bf v}=\frac{\partial {\bf x}}{\partial t}$ corresponding to the three cases in definition 1 are given by formulas (\ref{def1_eq1})-(\ref{def1_eq3}) where $\pmb{\chi}$ is replaced by $\displaystyle \frac{\partial \pmb{\chi}}{\partial t}$. In this paper we will work under these simplifying assumptions. Lastly, we notice that in the one dimensional case when $\alpha \neq 1$ is a constant, the fluid described by the fractional model shows shear thinning if $\alpha > 1$, or shear thickening if $\alpha<1$ \cite{dra2018}.

\subsection{Power Law Model}

The constitutive equation of an incompressible power-law fluid is \cite{cr}:
\begin{equation}
{\pmb{\sigma}}=-p\,{\bf I}+2\mu(II_{\bf D})\,{\bf D}, 
\label{eq9}
\end{equation}
where ${\pmb{\sigma}}$ and $p$ represent the Cauchy stress tensor and, respectively, the hydrostatic pressure, and ${\bf D}$ is the rate of deformation tensor of components:
\begin{equation}
\displaystyle D_{ij}=\frac{1}{2}\left( \frac{\partial v_{i}}{\partial x_{j}}+\frac{\partial v_{j}}{\partial x_{i}}\right),\, i,j=1,2,3
\label{eq10}
\end{equation}
Here ${\bf v}=(v_{1},\,v_{2},\,v_{3})$ is the velocity field from the classical theory of continuum mechanics. The second invariant of the second order tensor ${\bf D}$ is $II_{\bf D}=-{\rm tr}\left( {\bf D}^{2} \right)/2$ since the fluid is incompressible so ${\rm tr}({\bf D})=0$.  The viscosity of the power-law fluid is given by:
\begin{equation}
\mu(II_{\bf D})=K\,\left(-4II_{\bf D}\right)^{(n-1)/2}
\label{eq11}
\end{equation}
The two parameters of the model are $n$, a positive real number called the power law index, and $K$, the consistency (of the physical units). When $n=1$ the constitutive equation (\ref{eq9}) reduces to the classic equation of a viscous Newtonian fluid. When $n \neq 1$ the fluid shows non-Newtonian behavior: shear thinning if $n<1$, or shear thickening if $n>1$. 

\subsection{Casson's Model}
The constitutive equation of the Casson's model is \cite{fung}:
\begin{equation}
{\pmb{\sigma}}=-p\,{\bf I}+2\mu(J_{2}){\bf D},
\label{cass1}
\end{equation}
where ${\pmb{\sigma}},\;p,$ and ${\bf D}$ represent the Cauchy stress tensor, the hydrostatic pressure, and the rate of deformation tensor, respectively, and:
\begin{equation}
\mu(J_{2})=\left[ \left( \eta^{2}J_{2}\right)^{1/4}+\left(\frac{\tau_{y}}{2} \right)^{1/2} \right]^{2}\,J_{2}^{-1/2},
\label{cass_mu}
\end{equation}
with $J_{2}=\frac{1}{2}D_{ij}D_{ij}$ \footnote{Although $J_{2}=-II_{\bf D}$ we will use $J_{2}$ for the Casson's model and $II_{\bf D}$ for the power law fluid since these are the common notations used by these two models.} is the second invariant of ${\bf D}$. Parameter $\eta$ is called Casson's coefficient of viscosity, and $\tau_{y}$ is a constant yield stress in shear.  The following yielding condition provides information about the stress needed for the flow to happen:
\begin{equation}
D_{ij}=\left\{
\begin{array}{ll}
0, & J_{2}'<k\\
\frac{1}{2\mu}\sigma_{ij}', & J_{2}'\geq k
\end{array}
\right.
\label{cass2}
\end{equation}

for $J_{2}'=\frac{1}{2}\sigma_{ij}'\sigma_{ij}'$ and $\sigma_{ij}'=\sigma_{ij}-\frac{1}{3}\sigma_{kk}\delta_{ij}$. Equations (\ref{cass1}) and (\ref{cass2}) model the blood as a viscoplastic material.

\section{Poiseuille Flow}

In this section we consider the three-dimensional fully developed steady laminar flow of an incompressible non-Newtonian fluid through an horizontal circular pipe with rigid and impermeable walls. Let $R$ be the constant radius of the pipe. The flow is axi-symmetric and is driven by an externally imposed pressure gradient. For simplicity, body forces are neglected. In cylindrical coordinates $(r,\theta,z)$, the components of fluid's velocity are: 
\begin{equation}
v_{r}=0,\;v_{\theta}=0,\;v_{z}=w(r),
\label{vel}
\end{equation}
 and therefore the equation of continuity:
\begin{equation}
\displaystyle \frac{\partial v_{r}}{\partial r}+\frac{v_{r}}{r}+\frac{\partial v_{z}}{\partial z}=0,
\label{ec}
\end{equation}
 is identically satisfied. The equilibrium equations for the components of the Cauchy stress tensor $\pmb{\sigma}(r,z)$ in cylindrical coordinates are:
 \begin{align}
 \displaystyle \frac{\partial \sigma_{rr}}{\partial r}+\frac{1}{r} \frac{\partial \sigma_{r \theta}}{\partial \theta}+\frac{\partial \sigma_{rz}}{\partial z}+\frac{\sigma_{rr}-\sigma_{\theta \theta}}{r}&=0,\nonumber\\
 \displaystyle \frac{\partial \sigma_{r \theta}}{\partial r}+\frac{1}{r}\frac{\partial \sigma_{\theta \theta}}{\partial \theta}+
 \frac{\partial \sigma_{\theta z}}{\partial z}+\frac{2}{r}\sigma_{r \theta}&=0,\nonumber\\
 \displaystyle \frac{\partial \sigma_{rz}}{\partial r}+\frac{1}{r}\frac{\partial \sigma_{\theta z}}{\partial \theta}+\frac{\partial \sigma_{zz}}{\partial z}+\frac{1}{r}\sigma_{rz}&=0.
 \label{ee}
 \end{align}
The no slip boundary condition at $r=R$ is:
\begin{equation}
w(R)=0
\label{bc1}
\end{equation}
and the boundary condition at $r=0$ expressing the axial symmetry of the flow will be specified for each constitutive model separately. \\
We further solve the above boundary value problem for the three constitutive laws reviewed in the previous section.
 
 \subsection{Fractional Model}
 The only non-zero components of $\pmb{\sigma}$ given by (\ref{eq3}) are:
 \begin{align}
 \sigma_{rr}&=\sigma_{\theta \theta}=\sigma_{zz}=-p(r,z), \label{claw_p} \\
 \sigma_{rz}&=\sigma_{zr}=\displaystyle \mu \frac{1}{\Gamma(m-\alpha)}\frac{d^{m}}{dr^{m}}\int_{0}^{r} \frac{1}{(r-\tau)^{\alpha+1-m}}\left( w(\tau) - \sum_{k=0}^{m-1}\frac{\tau^{k}}{k!}\frac{d^{k}}{d\tau^{k}}w(0^+)\right)d\tau\nonumber\\
 &=\mu D_{r}^{\alpha}w(r) \label{claw_shear}
 \end{align}
Here the region of influece is $H=[0,r]$. We assume that  $\displaystyle \frac{d^{m}w}{dr^{m}}\in L^{1}([0,R])$, and introduce an extra term under the integral in (\ref{claw_shear}) which is not present in definition 1 so that we can impose non-zero boundary conditions. In this case the integral in (\ref{claw_shear}) is the left-sided Caputo fractional derivative of order $\alpha$:
 \begin{equation}
 D_{r}^{\alpha}w(r)=\begin{cases} J^{m-\alpha}\frac{d^{m}w(r)}{dr^{m}}= \displaystyle \frac{1}{\Gamma(m-\alpha)}\int_{0}^{r} \frac{1}{(r-\tau)^{\alpha+1-m}}\frac{d^{m}w(\tau)}{d\tau^{m}}d\tau, & m-1<\alpha < m; \\
 \frac{d^{m}w(r)}{dr^{m}}, & \alpha = m;
 \end{cases}
 \label{caputo}
 \end{equation} 
for $m\in \{1,2,3,...\}$.
We notice that the definition of the Riemann fractional integral operator of order $\alpha$ denoted by $J^{\alpha}$ was also given in formula (\ref{caputo}). \\  

As in \cite{dra2018}, we assume the following boundary condition at $r=0$:
\begin{equation} 
\frac{d^{k}}{dr^{k}}w(0^+)=0,\,k=1,2,...m-1
\label{zero}
\end{equation}
Assume further that the two parameters of the fractional model, $\mu$ and $\alpha$, are constants. By replacing equations (\ref{claw_p}) and (\ref{claw_shear}) into the system of equations (\ref{ee}) gives $p=p(z)$ and:
\begin{equation}
\displaystyle \mu \frac{1}{r}\frac{d}{dr}\left( r\,D_{r}^{\alpha}w(r)\right)=\frac{dp}{dz}
\label{shear_eq}
\end{equation}
Here $\frac{dp}{dz}=C<0$ is a known constant. Integrating equation (\ref{shear_eq}) and using the boundary conditions (\ref{bc1}) and (\ref{zero}), and properties of Caputo fractional integrals and derivatives give \cite{dra2018}:
\begin{equation}
\displaystyle w(r)=-\frac{C}{2\mu\alpha (\alpha+1)\Gamma(\alpha)}\left( R^{\alpha+1} - r^{\alpha+1} \right).
\label{sol_wFF}
\end{equation}
The volumetric flow rate is then:
\begin{equation}
\displaystyle Q = \int_{0}^{R} 2\pi\,r w(r)dr=-\frac{\pi C}{2\mu \alpha (\alpha+3) \Gamma(\alpha)} R^{\alpha+3}
\label{Q1}
\end{equation}
and the average velocity in this case is:
\begin{equation}
W_{ave}=\displaystyle \frac{Q}{\pi R^{2}}=-\frac{C}{2\mu \alpha (\alpha+3) \Gamma(\alpha)} R^{\alpha+1}
\label{ave_Wfrac}
\end{equation}

\subsection{Power Law Model}
The only non-zero components of ${\pmb{\sigma}}$ given by (\ref{eq9}) and (\ref{eq11}) are:
\begin{align}
\sigma_{rr}&=\sigma_{\theta \theta}=\sigma_{zz}=-p(r,z), \label{law_p}\\
 \sigma_{rz}&=\sigma_{zr}=\displaystyle K \left( -\frac{dw}{dr} \right)^{n}
 \label{law_shear}
 \end{align}
where we used the fact that $w$ decreases with increasing $r$ and that $\sigma_{rz}>0$. The boundary condition at $r=0$ is:
 \begin{equation}
 \displaystyle \frac{dw}{dr}(0)=0.
 \label{bc0}
 \end{equation}
 By replacing formulas (\ref{law_p}) and (\ref{law_shear}) into equations (\ref{ee}) we obtain again that $p=p(z)$. As before, we denote by $C=\frac{dp}{dz}<0$ a known constant. If we integrate the third equation of system (\ref{ee}) and use the boundary conditions (\ref{bc1}) and (\ref{bc0}) we get \cite{cr}:
\begin{equation}
\displaystyle w(r)=\frac{n}{n+1}\left( -\frac{C}{2K} \right)^{1/n}\left( R^{1+1/n}-r^{1+1/n}\right).
\label{sol_wPF}
\end{equation}
The volumetric flow rate is now:
\begin{equation}
\displaystyle Q = \int_{0}^{R} 2\pi\,r w(r)dr=\frac{\pi n}{3n+1}\left( -\frac{C}{2K} \right) ^{1/n} R^{3+1/n}
\label{Q2}
\end{equation}
and the average velocity is:
\begin{equation}
W_{ave}=\displaystyle \frac{Q}{\pi R^{2}}=\frac{n}{3n+1}\left( -\frac{C}{2K} \right) ^{1/n} R^{1+1/n}
\label{ave_Wpow}
\end{equation}
The two parameters of the power law model are $n$ and $K$.

\subsection{Casson's Model}
The only non-zero components of ${\pmb{\sigma}}$ given by (\ref{cass1}) and (\ref{cass_mu}) are:
\begin{align}
\sigma_{rr}&=\sigma_{\theta \theta}=\sigma_{zz}=-p(r,z), \label{cass_p}\\
 \sigma_{rz}&=\sigma_{zr}=\displaystyle -\left( \sqrt{\eta}\sqrt{-\frac{dw}{dr}}+\sqrt{\tau_{y}}\right)^{2}
 \label{cass_shear}
 \end{align}
where we used again the fact that $w$ decreases with increasing $r$ (so $\frac{dw}{dr}<0$) and that $\sigma_{rz}>0$. By replacing (\ref{cass_p}) and (\ref{cass_shear}) into equations (\ref{ee}) we obtain $p=p(z)$ and if we let $C=\frac{dp}{dz}<0$ be a known constant we can integrate once the third equation of system (\ref{ee}) and use the boundness of the shear stress at $r=0$ to obtain:
\begin{equation}
\sigma_{rz}=-\frac{C}{2}r
\label{shear}
\end{equation}
Let $r_{c}$ be a characteristic radius such that the yielding shear stress can be written from (\ref{shear}) as:
\begin{equation}
\tau_{y}=-\frac{C}{2}r_{c}
\label{yield}
\end{equation}
The yielding condition says that the blood will not flow if $0 \leq r < r_{c}$ and will flow if $r_{c}\leq r \leq R$.  
If we now replace formula (\ref{cass_shear}) into (\ref{shear}), integrate on $[r_{c},R]$ and use the no slip boundary condition (\ref{bc1}) then the following expression is obtained for the speed $w(r)$ \cite{fung}:
\begin{equation}
\displaystyle w(r)=\frac{1}{\eta}\left(\frac{C}{4}(r^{2}-R^{2})+\frac{4}{3\sqrt{2}}\sqrt{-C\tau_{y}}(r^{3/2}-R^{3/2})-\tau_{y}
(r-R)\right),\; r_{c}\leq r \leq R
\label{cass_w1}
\end{equation}
If we replace formula (\ref{yield}) into formula (\ref{cass_w1}) and calculate the core speed $w_{c}$ for $r=r_{c}$ we finally obtain:
\begin{equation}
\displaystyle w(r) = \begin{cases}
-\frac{C}{4\eta}\left( R^{2}-r^{2}-\frac{8}{3}\sqrt{r_{c}}(R^{3/2}-r^{3/2})+2r_{c}(R-r)\right), & r_{c} \leq r \leq R\\
-\frac{C}{4\eta}(\sqrt{R}-\sqrt{r_{c}})^{3}(\sqrt{R}+\frac{1}{3}\sqrt{r_{c}}), & 0\leq r \leq r_{c}
\end{cases}
\label{cass_w}
\end{equation} 
For $r_{c}\leq R$ the volumetric flow rate is:
\begin{equation}
\displaystyle Q = \int_{0}^{R} 2\pi\,r w(r)dr=-\frac{\pi C R^{4}}{8\eta}\left(1-\frac{16}{7}\left( \frac{r_{c}}{R} \right)^{1/2}
+\frac{4}{3}\left( \frac{r_{c}}{R} \right) -\frac{1}{21}\left( \frac{r_{c}}{R} \right)^{4} \right)
\label{Q3}
\end{equation}
and thus the average velocity is:
\begin{equation}
W_{ave}=\displaystyle \frac{Q}{\pi R^{2}}=-\frac{C R^{2}}{8\eta}\left(1-\frac{16}{7}\left( \frac{r_{c}}{R} \right)^{1/2}
+\frac{4}{3}\left( \frac{r_{c}}{R} \right) -\frac{1}{21}\left( \frac{r_{c}}{R} \right)^{4} \right)
\label{ave_Wcass}
\end{equation}
 The two parameters of the Casson's model are $\eta$ and $r_{c}$.

\section{Results}

\begin{figure}[htbp]
\centerline{\includegraphics[clip, trim=1cm 8.5cm 1cm 9cm, width=\textwidth]{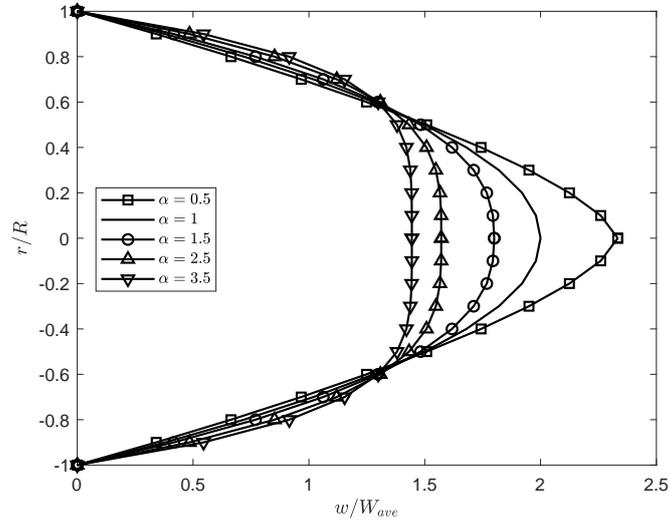}}
\caption{Non-dimensional speeds $w(r)/W_{ave}$ for the fractional model calculated using formulas (\ref{sol_wFF}) and (\ref{ave_Wfrac}) for various values of the fractional order $\alpha$. When $\alpha=1$ the speed profile corresponds to the incompressible viscous Newtonian fluid.}
\label{fig1}
\end{figure}

\begin{figure}[htbp]
\centerline{\includegraphics[clip, trim=1cm 8.5cm 1cm 9cm, width=\textwidth]{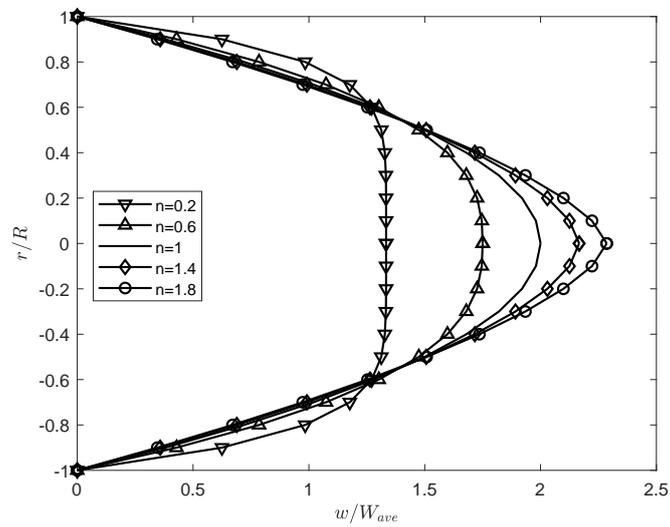}}
\caption{Non-dimensional speeds $w(r)/W_{ave}$ for the power law model calculated using formulas (\ref{sol_wPF}) and (\ref{ave_Wpow}) for various values of the power index $n$. When $n=1$ the speed profile corresponds to the incompressible viscous Newtonian fluid.}
\label{fig2}
\end{figure}

\begin{figure}[htbp]
\centerline{\includegraphics[clip, trim=1cm 8.5cm 1cm 9cm, width=\textwidth]{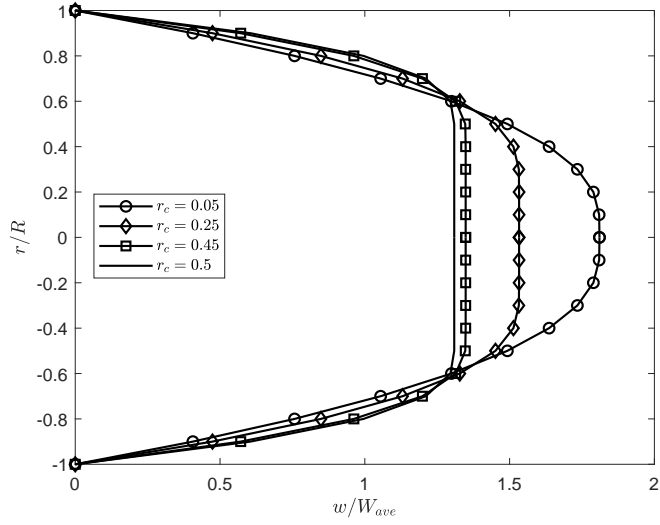}}
\caption{Non-dimensional speeds $w(r)/W_{ave}$ for the Casson's model calculated using formulas (\ref{cass_w}) and (\ref{ave_Wcass}) for various values of $r_{c}$.}
\label{fig3}
\end{figure}

We start this section with plots of the dimensionless speeds $w/W_{ave}$ calculated for each model. For simplicity, in our simulations we use the following values: 
\begin{align}
\mu&=1\,g/(cm^{2-\alpha}\times s),\; K=1\,g/(cm\times s^{2-n}),\;\eta=1\,g/(cm \times s), \nonumber\\
R&=1\,cm,\;C=-1\,g/(cm^{2}\times s^{2})
\label{consts}
\end{align}
Figures \ref{fig1}-\ref{fig3} show dimensionless speed profiles predicted by the three models for various values of the corresponding second parameter which was not fixed in (\ref{consts}). We observe that the bluntness of the velocity profiles increases with increasing $\alpha$ for the fractional model (fig. \ref{fig1}), with decreasing $n$ for the power law model (fig. \ref{fig2}), and with increasing $r_{c}$ for the Casson's model (fig. \ref{fig3}). Although the velocity profiles of the three models look similar, the physical meanings of the varying parameters $\alpha$, $n$ and $r_{c}$ are different. The fractional order $\alpha \neq 1$ of the fractional model is a measure of the non-local interactions of blood's particles during flow, the power law index $n \neq 1$ of the power law model determines the local dependence of the apparent viscosity on the shear rate, and $r_{c}$ of the Casson's model is the radius at which yield stress is attained and the blood starts to flow in the region
 $r_{c}\leq r \leq R$. 

\begin{figure}[htbp]
\centerline{\includegraphics[clip, trim=1cm 8.5cm 1cm 9cm, width=\textwidth]{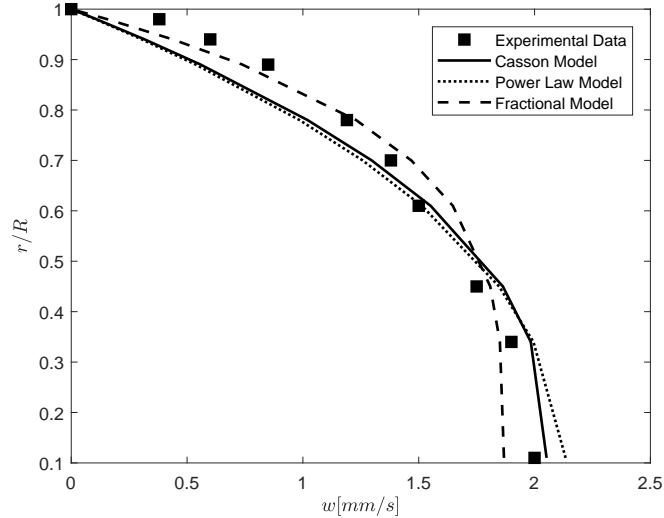}}
\caption{Curve fitting to experiments performed on blood {\it in vitro}. The experimental data are taken from \cite{srir} which is a subset of the data published in \cite{long}.}
\label{fig4}
\end{figure}

\begin{figure}[htbp]
\centerline{\includegraphics[clip, trim=1cm 8.5cm 1cm 9cm, width=\textwidth]{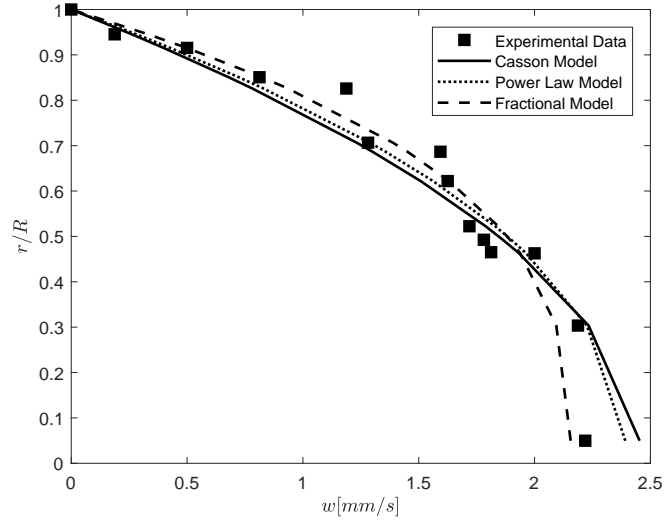}}
\caption{Curve fitting to experiments performed on blood {\it in vivo}. The experimental data are taken from \cite{long}.}
\label{fig5}
\end{figure}

In figs. \ref{fig4} and \ref{fig5} we present curve fitting results to experimental data given in \cite{long}. The fitting was performed in Matlab using the built-in function {\bf lsqcurvefit} that solves non-linear least squares problems using the trust region reflective and Levenberg-Marquardt algorithms. The {\it in vitro} experimental data used in fig. \ref{fig4} are a subset (the same as in \cite{srir}) of the data obtained from a glass tube of radius $R=27.1\,\mu m$ steadily perfused with human blood at a measured pressure gradient $-C=3732\,dyn/cm^{3}$. The following values for the best fits were obtained: 
\[
\alpha=3.32,\,\mu = 0.39\,g/(\mu m^{2-\alpha} \times s)
\]
for the fractional model,
\[
n=0.67,\;K=0.0015\,g/(\mu m \times s^{2-n})
\]
for the power law model, and
\[
r_{c}=2.98\,\mu m,\; \eta=0.0011\,g/(\mu m \times s),
\]
for the Casson's model. The Euclidean norm of the difference between each fit curve and the experimental data is 0.37 for the fractional model, 0.59 for the power law model, and 0.54 for the Casson's model. \\

The {\it in vivo} experimental data used in fig. \ref{fig5} were obtained from blood flow through a venule of radius $R=20.1\,\mu m$ of a mouse cremaster muscle before systemic hemodilution \cite{long}. Since the pressure gradient was not measured, we used the reported estimations given in \cite{long} to approximate  $-C=7122\,dyn/cm^{3}$. The following values for the best fits were obtained:
\[
\alpha=1.97,\,\mu = 0.021\,g/(\mu m^{2-\alpha} \times s)
\]
for the fractional model,
\[
n=0.81,\;K=0.0021\,g/(\mu m \times s^{2-n})
\]
for the power law model, and
\[
r_{c}=1.97*10^{-6}\,\mu m,\; \eta=0.0029\,g/(\mu m \times s),
\]
for the Casson's model. The Euclidean norm of the difference between each fit curve and the experimental data is 0.41 for the fractional model, 0.53 for the power law model, and 0.62 for the Casson's model. \\

The Euclidean norms of the fractional model are smaller than the others for both sets of experimental data, suggesting that the fractional model might provide a better fit. Lastly, we notice that the power law model was sensitive to the initial guess required by Matlab's built-in function {\bf lsqcurvefit}, while the fractional and Casson's models were robust to changes in these initial guesses.

\section{Conclusions and Future Work}

In this paper we presented a comparison of three models of blood flow: a fractional model, the power law model, and the Casson's model. We presented analytic solutions for the Poiseuille flow through an axi-symmetric  circular tube with rigid and impermeable walls and no-slip boundary condition at the fluid-wall interface. The flow is due to a pressure gradient applied along the tube's axis. Although the velocity profiles of the three models look alike, the physical assumptions are different. The fractional model is non-local in nature, combining the arrangements of particles and motion. The power law and Casson's models abelong to the class of non-Newtonian viscosity models since the shear stress is expressed as a product between an apparent viscosity that depends on the flow's mechanical invariants and the rate of deformation tensor. These two models are local in nature because they use the classical rate of deformation tensor. In addition, Casson's model uses a yielding condition to describe blood's viscoplasticity.  Lastly, best fit curves to experimental data from \cite{long} indicate that the fractional model provides the better fit among the three models. In our future work we plan to use the fractional model in studies of cerebral blood flow where the fractional order $\alpha$ could play the role of a biomarker for early detection of disease.


\section*{\bf REFERENCES}

\begingroup
\renewcommand{\section}[2]{}%

\endgroup


\begin{thebibliography}{9}

\bibitem{albu1}
I.D.~Albu, M.~Neamtu and D.~Opris,
\textit{The geometry of fractional osculator bundle of higher order and applications}, 
Anal. St. Univ. Al.I.Cuza Iasi, Proceedings of the Conf. on Diff. Geometry: Lagrange and Hamilton spaces,
(2007), 
21–-31. 

\bibitem{albu2}
I.D.~Albu and D.~Opris,
\textit{The geometry of fractional tangent bundle and applications}, 
BSG Proceedings 16. The Int. Conf. of Diff. Geom. and Dynamical Systems (DGDS-2008) and The V-th Int. Colloq. of Mathematics in Engineering and Numerical Physics (MENP-5),
(2008),
1--11.

\bibitem{apostol}
A.J.~Apostolidis and A.N.~Beris,
\textit{Modeling of the blood rheology in steady-state shear flows},
J. Rheol.,
\textbf{58} (3) (2014),
 607--633.
 
 \bibitem{bal}
D.~Baleanu, K.~Diethelm, E.~Scalas and J.J.~Trujillo,
\textit{Fractional Calculus: Models and Numerical Methods}, 
World Scientiﬁc, Hackensack, NJ, 2012.

\bibitem{bess}
N.~Bessonov, A.~Sequeira, S.~Simakov, Y.~Vassilevskii and V.~Volpert,
\textit{Methods of blood flow modelling},
Math. Model. Nat. Phenom.
\textbf{11} (1) (2016),
1--25.

\bibitem{carp1}
A.~Carpinteri and P.~ Cornetti, 
\textit{A fractional calculus approach to the description of stress and strain localization in fractal media}, 
Chaos Solitons Fractals, 
\textbf{13} (2002),
85–-94.

\bibitem{carp2}
A.~Carpinteri, P.~Cornetti and A.~Sapora, 
\textit{Static-kinematic fractional operators for fractal and non-local solids}, 
Z. Angew. Math. Mech., 
\textbf{89}(3) (2009),
207–-217.

\bibitem{cr}
R.P.~Chhabra and J.F.~Richardson,
\textit{Non-Newtonian Flow and Applied Rheology},
Butterworth-Heinemann, Woburn, MA, 2008.
 
\bibitem{cott} 
M.~Cottone, M.~Di Paola and M.~Zingales, 
\textit{Fractional mechanical model for the dynamics of non-local continuum}, 
Advances in Numerical Methods,
(2009),
389–-423.

\bibitem{cot} 
K.~Cottrill-Sheperd and M.~Naber,
\textit{Fractional differential forms}, 
J.Math.Phys., 
\textbf{42} (2001), 
2203–-2212.

\bibitem{dav}
P.F.~Davies,
\textit{Hemodynamic shear stress and the endothelium in cardiovascular pathophysiology},
Nat. Clin. Pract. Cardiovasc. Med.
\textbf{6} (1) (2009),
16--26.

\bibitem{di1} 
M.~DiPaola and M.~Zingales, 
\textit{Long-range cohesive interactions of non-local continuum faced by fractional calculus}, 
Int. J. Solids Struct., 
\textbf{45} (2008), 
5642–-5659.

\bibitem{di2}
M.~DiPaola, A.~Pirrotta and M.~Zingales, 
\textit{Mechanically-based approach to non-local elasticity: variational principles}, 
Int. J. Solids Struct., 
\textbf{47} (5) (2010),
539–-548.

\bibitem{dra_siv}
C.S.~Drapaca and S.~Sivaloganathan, 
\textit{A fractional model of continuum mechanics}, 
J. Elast.,
\textbf{107} (2012),
105–-123.

\bibitem{dra2018}
C.S.~Drapaca,
\textit{Poiseuille flow of a non-local non-Newtonian fluid with wall slip: a first step in modeling cerebral microaneurysms},
(2018),
arXiv:1801.04917 [physics.flu-dyn].

\bibitem{fung}
Y.C.~Fung,
\textit{Biomechanics: Mechanical Properties of Living Tissues},
Springer-Verlag New York, Inc. New York, NY, 1993.

\bibitem{hilfer}  
R.~Hilfer, 
\textit{Applications of Fractional Calculus in Physics},
World Scientiﬁc, River Edge, NJ, 2000.

\bibitem{iad}
C.~Iadecola and M.~Nedergaard,
\textit{Glial regulation of the cerebral microvasculature},
Nature Neuroscience,
\textbf{10} (11) (2007),
1369--1376.

\bibitem{john}
B.M.~Johnston, P.R.~Johnston, S.~Corney and D.~Kilpatrick,
\textit{Non-Newtonian blood flow in human right coronary arteries: steady state simulations},
Journal of Biomechanics, 
\textbf{37} (2004),
709–-720.

\bibitem{ku}
D.N.~Ku,
\textit{Blood flow in arteries},
Annu. Rev. Fluid Mech.
\textbf{29} (1997),
399--434.

\bibitem{leibniz}
G.~Leibniz, 
\textit{Letter to L’Hospital}, 
(1695).

\bibitem{liouville} 
J.~Liouville, 
\textit{Sur le calcul des differentielles a indices quelconques}, 
J. \'{E}c. Polytech., 
\textbf{13} (1832), 
71.

\bibitem{long}
D.S.~Long, M.L.~Smith, A.R.~Pries, K.~Ley and E.R.~Damiano, 
\textit{Microviscometry reveals reduced blood viscosity and altered shear rate and shear stress profiles in microvessels after hemodilution}, Proc. Natl. Acad. Sci. USA
\textbf{101} (2004),
10060–10065. 

\bibitem{mart}
N.S.~Martys, W.L.~George, B.-W.~Chun and D.~Lootens,
\textit{A smoothed particle hydrodynamics-based fluid model with a spatially dependent viscosity: application to flow of a suspension with a non-Newtonian fluid matrix},
Rheologica Acta,
\textbf{49} (10) (2010),
1059--1069.

\bibitem{mili}
C.~Milici and G.~Draganescu, 
\textit{New Methods and Problems in Fractional Calculus},
LAP LAMBERT Academic Publishing, 
Saarbrucken, Germany, 2015.

\bibitem{nadim}
S.~Nadeem, N.S.~Akbar, A.A.~Hendi and T.~Hayat,
\textit{Power law fluid model for blood flow through a tapered artery with a stenosis},
Applied Mathematics and Computation, 
\textbf{217} (2011),
7108--7116.

\bibitem{old_spa}
K.B.~Oldham and J.~Spanier, 
\textit{The Fractional Calculus: Theory and Applications of Differentiation and Integration to Arbitrary Order}, 
Dover,
Mineola, NY, 2006.

\bibitem{os}
M.~Ostoja-Starzewski, J.~Li, H.~Joumaa and P.N.~Demmie, 
\textit{From fractal media to continuum mechanics}, 
Z. Angew. Math. Mech.,
\textbf{94} (5) (2014),
373–-401.

\bibitem{park}
H.~Park, E.~Yeom, S.-J.~Seo, J.-H.~Lim and S.-J.~Lee,
\textit{Measurement of real pulsatile blood flow using X-ray PIV technique with CO2 microbubbles},
Scientific Reports,
\textbf{5} (2015),
8840.

\bibitem{pod}
I.~Podlubny, 
\textit{Fractional Differential Equations},
Academic Press, 
San Diego, California, 1999.

\bibitem{pod1}
I.~Podlubny, 
\textit{Geometrical and physical interpretation of fractional integration and fractional differentiation}, 
Fractional Calc.Appl.Anal., 
\textbf{5} (4) (2002),
367–-386.

\bibitem{pop}
A.S.~Popel and P.C.~Johnson,
\textit{Microcirculation and hemorheology},
Annu. Rev. Fluid Mech.
\textbf{37} (2005),
43--69.

\bibitem{pri}
A.R.~Pries, D.~ Neuhaus and P.~Gaehtgens,
\textit{Blood viscosity in tube flow: dependence on diameter and hematocrit},
American Journal of Physiology - Heart and Circulatory Physiology,
\textbf{263} (6) (1992),
H1770--H1778.

\bibitem{riemann} 
B.~Riemann, 
\textit{Versuch einer allgemeinen Auffassung der Integration and Differentiation}, 
Gesammelte Werke,
(1876),
62.

\bibitem{samko}
S.G.~Samko, A.A. ~Kilbas and O.I.~Marichev, 
\textit{Fractional Integrals and Derivatives},
Gordon and Breach, 
Yverdon, Switzerland, 1993.

\bibitem{santi}
T.P.~Santisakultarm, N.R.~Cornelius, N.~Nishimura, A.I.~Schafer, R.T.~Silver, P.C.~Doerschuk, W.L.~Olbricht and C.B.~Schaffer,
\textit{In vivo two-photon excited fluorescence microscopy reveals cardiac- and respiration-dependent pulsatile blood flow in cortical blood vessels in mice},
Am J Physiol Heart Circ Physiol.,
\textbf{302} (7) (2012), 
H1367--H1377.

\bibitem{sou}
J.V.~Soulis, G.D.~Giannoglou, Y.S.~Chatzizisis, K.V.~Seralidou, G.E.~Parcharidis and G.E.~Louridas,
\textit{Non-Newtonian models for molecular viscosity and wall shear stress in a 3D reconstructed human left coronary artery},
Medical Engineering $\&$ Physics,
\textbf{30} (2008), 
9–-19.

\bibitem{srir}
K.~Sriram, M.~Intaglietta and D.M.~Tartakovsky,
\textit{Non-Newtonian flow of blood in artetioles: consequences for wall shear stress measurements},
Microcirculation,
\textbf{21}(7) (2014),
628--639.

\bibitem{sum1}
W. Sumelka,
\textit{Fractional viscoplasticity}, 
Mech. Research Communications, 
\textbf{56} (2014), 
31-36.

\bibitem{sum2}
W. Sumelka and T. Blaszczyk, 
\textit{Fractional continua for linear elasticity},
Arch. Mech. 
\textbf{66} (3)(2014), 
147-172.

\bibitem{sum3}
W. Sumelka, 
\textit{On fractional non-local bodies with variable length scale},
Mech. Research Communications,
\textbf{86} (2017), 
5-10.

\bibitem{tar}
 V.E.~Tarasov, 
\textit{Fractional Dynamics: Applications of Fractional Calculus to Dynamics of Particles, Fields and Media},
Springer, 
Heidelberg, Germany, 2010.

\bibitem{west}
B.J.~West, 
\textit{Fractional Calculus View of Complexity: Tomorrow’s Science},
CRC Press, 
Boca Raton, FL, 2015.

\bibitem{yil}
G.~Yilmaz and M.Y.~Gundogdu,
\textit{A critical review on blood flow in large arteries; relevance to blood rheology, viscocity models, and physiologic conditions},
Korea-Australia Rheology Journal,
\textbf{20} (4) (2008),
197--211.

\bibitem{zhong}
Z.~Zhong, H.~Song, T.Y. P.~Chui, B.L.~Petrig and S.A.~Burns,
\textit{Noninvasive measurements and analysis of blood velocity profiles in human retinal vessels},
Investigative Ophthalmology $\&$ Visual Science,
\textbf{52} (2011), 4151--4157.
 
\end{thebibliography}
\end{document}